\documentclass[twocolumn,showpacs,showkeys,amsmath,amssymb,superscriptaddress,floatfix,nofootinbib]{revtex4}

\usepackage{graphicx}
\usepackage{bm}
\usepackage{subfigure}

\makeatletter
\newcommand\erfc{\mathop{\operator@font erfc}\nolimits}
\def\slashchar#1{\setbox0=\hbox{$#1$}
   \dimen0=\wd0 \setbox1=\hbox{/} \dimen1=\wd1
   \ifdim\dimen0>\dimen1 \rlap{\hbox to \dimen0{\hfil/\hfil}} #1
   \else  \rlap{\hbox to \dimen1{\hfil$#1$\hfil}} / \fi}

\begin{document}

\title{
Explanation of hadron transverse-momentum spectra in heavy-ion collisions at \mbox{$\sqrt{s_{\rm NN}}$ = 2.76 TeV} within chemical non-equilibrium statistical hadronization model
}

\author{Viktor Begun}
 \email{viktor.begun@gmail.com}
 \affiliation{Institute of Physics, Jan Kochanowski University, PL-25406~Kielce, Poland}
 \affiliation{Bogolyubov Institute for Theoretical Physics, 03680 Kiev, Ukraine}

\author{Wojciech Florkowski}
\email{wojciech.florkowski@ifj.edu.pl}
\affiliation{Institute of Physics, Jan Kochanowski University, PL-25406~Kielce, Poland}
\affiliation{The H. Niewodnicza\'nski Institute of Nuclear Physics, Polish Academy of Sciences, PL-31342 Krak\'ow, Poland}

\author{Maciej Rybczynski}
\email{maciej.rybczynski@ujk.edu.pl}
\affiliation{Institute of Physics, Jan Kochanowski University, PL-25406~Kielce, Poland}

\date{May 30, 2014}

\begin{abstract}
A chemical non-equilibrium version of the statistical hadronization model combined with a single-freeze-out scenario is used to analyze the transverse-momentum spectra of pions, kaons and protons produced in heavy-ion collisions at $\sqrt{s_{\rm NN}}$ = 2.76 TeV. With the statistical parameters found in earlier studies of hadronic ratios and two new geometric parameters determined by the absolute normalization and slopes of the spectra, we obtain a very good agreement between the data and the model predictions. In particular,  the enhancement of pions observed at the very low transverse-momenta is very well reproduced. In the chemical non-equilibrium approach, this effect may be naturally interpreted  as the onset of pion condensation in heavy-ion collisions. We further stress that the chemical non-equilibrium explains not only the proton yield but the proton spectra up to $p_T=$~3 GeV/c as well.
\end{abstract}



\maketitle

Statistical models of hadron production \cite{Koch:1985hk,Cleymans:1992zc,
Cleymans:1998fq,Gazdzicki:1998vd,
Braun-Munzinger:1994xr,
Cleymans:1996cd,Becattini:2000jw,
Sollfrank:1993wn,Schnedermann:1993ws,
Braun-Munzinger:1995bp,Becattini:1997uf,
Yen:1998pa,Braun-Munzinger:1999qy,Becattini:2003wp,
Braun-Munzinger:2001ip,Florkowski:2001fp,
Broniowski:2001we,Broniowski:2001uk,
Retiere:2003kf} have become one of the cornerstones of our understanding of ultra-relativistic heavy-ion collisions. They have explained successfully the data on hadronic abundances collected at the AGS \cite{Braun-Munzinger:1994xr,Cleymans:1996cd,
Becattini:2000jw}, SPS \cite{Sollfrank:1993wn,Schnedermann:1993ws,
Braun-Munzinger:1995bp,Becattini:1997uf,
Yen:1998pa,Braun-Munzinger:1999qy,Becattini:2003wp}, and RHIC energies \cite{Braun-Munzinger:2001ip,Florkowski:2001fp,
Broniowski:2001we,Broniowski:2001uk,
Retiere:2003kf}. Moreover, with the appropriate definition of the freeze-out space-time geometry and flow, they have also been used to describe successfully the transverse-momentum spectra of the produced hadrons and other soft-hadronic observables \cite{Broniowski:2001we,Retiere:2003kf}.

With these successes in mind, the new LHC data collected in Pb+Pb
collisions at $\sqrt{s_{\rm NN}}$ = 2.76 TeV comes as a big surprise --- the measured proton abundances \cite{Abelev:2012wca,Abelev:2013vea}
do not agree with the most common version of the thermal model
based on the grand-canonical ensemble
\cite{Kalweit:2012zz,Stachel:2013zma}. The possible explanations
of this anomaly refer to the inclusion of the hadronic
rescattering in the final stage \cite{Becattini:2012xb} or to the
concept of the hadronization and subsequent freeze-out that take
place off chemical equilibrium
\cite{Petran:2013qla,Petran:2013lja}.

Besides the proton anomaly, the same LHC data exhibits another interesting feature --- the low-transverse-momentum pion spectra show enhancement by about 25\%--50\% with respect to the predictions
of various thermal and hydrodynamic models. See, for example, the lower panel of Fig.~1 in \cite{Abelev:2012wca}, Fig.~12~(a) in \cite{Abelev:2013vea}, the left lower
panel of Fig.~2 in \cite{Song:2013qma}, and Fig.~13 in \cite{Pang:2013pma}. We observe repeatedly a very characteristic convex shape of the ratios data/model and a substantial slope at which the data/model ratios rise with decreasing $p_T$. We emphasize that this behavior is in contrast with the measurements done at lower energies, for example at RHIC, where the low-$p_T$ pion spectra were explained very well by the combined effects of the flow and resonance decays \cite{Florkowski:2001fp,Broniowski:2001we,
Prorok:2006ve}.

In this Letter, we connect the proton anomaly with the pion enhancement effect and show that the two problems may be solved naturally within the statistical model which assumes chemical
non-equilibrium at the freeze-out stage. The physics picture behind the non-equilibrium model is a sudden hadronization of the quark-gluon plasma (QGP) similar to the condensation of supercooled water. For a long time, such a picture has been advocated by Rafelski and his collaborators in their studies of hadronic abundances \cite{Rafelski:2011ek}. At the RHIC and lower energies this approach was an alternative to other versions of the statistical
model, giving substantially lower freeze-out temperatures. However, no clear advantage of the chemical non-equilibrium approach has been established in this energy range. At the LHC energy, the chemical non-equilibrium approach seems to be supported by the present data, see Ref.~\cite{Floris}. In this work, we analyze the new  LHC data describing the $p_T$-spectra of pions, kaons, and protons. We combine the concept of chemical non-equilibrium with the Cracow single-freeze-out model \cite{Broniowski:2001we} used in the Monte-Carlo version as implemented in THERMINATOR \cite{Kisiel:2005hn,Chojnacki:2011hb}. Our results indicate that the concept of chemical non-equilibrium is an attractive scenario for the  hadronization process at the LHC as it describes  the high-precision hadron spectra in addition to the hadron yields.   

Besides the thermodynamic parameters which we take directly from \cite{Petran:2013qla,Petran:2013lja}, our model has only two extra parameters: the transverse size, $r_{\rm max}$, and the (invariant) time, $\tau_f$, which characterize the hadronic system at freeze-out. We fit these two parameters to the measured transverse-momentum spectra of pions and kaons and find a remarkable agreement~\footnote{Since the proton yield cannot be reproduced in the equilibrium model, we omit the protons in the fitting procedure. On the other hand, the pion and kaon yields may be well reproduced in the two versions of the thermal models, hence, it is desired to compare directly the pion and kaon spectra in the two frameworks.}. 
We note that the shape of spectra is described with only one parameter, $r_{\rm max}/\tau_f$, because the combination $V = \pi \tau_f r_{\rm max}^2$ fixes the overall normalization.

 In particular, we reproduce very well the low-$p_T$ region of the pion spectrum. This is so, since the chemical non-equilibrium model predicts the freeze-out conditions which are very close to the pion condensation point. Although the protons are not included in the fit, we find that the chemical non-equilibrium model explains well their spectrum, in addition to their yield.

Below, we present our calculations done with the chemical non-equilibrium statistical hadronization model (denoted  below as NEQ SHM) comparing the results with the equilibrium version (denoted  below as EQ SHM). In the two cases the hadron rapidity and transverse-momentum distributions are calculated from the Cooper-Frye formula
\begin{eqnarray}
\nonumber \\
\frac{dN}{dy d^2p_T} &=& \int d\Sigma_\mu
p^\mu f(p\cdot u),
\label{frye-cooper}
\end{eqnarray}
where $d\Sigma_\mu$ is an element of the freeze-out hypersurface and $u^\mu$ is the hydrodynamic flow at freeze-out. The distribution function $f(p\cdot u)$ consists of primordial (directly produced) and secondary (produced by resonance decays) contributions.

The primordial distribution of the $i$th hadron in the local rest frame, where $u^\mu=(1,0,0,0)$, has the form \cite{Torrieri:2004zz}
\begin{eqnarray}
f_i = \frac{g_i}{\Upsilon^{-1}_i
\exp(\sqrt{p^2+m^2_i} / T ) \mp 1}.
\label{share-ni} \\ \nonumber
\end{eqnarray}
Here the $-1 \,\,(+1)$ sign corresponds to bosons (fermions) and $g_i$ is the degeneracy factor connected with spin. The fugacity factor $\Upsilon_i$ is defined through the parameters $\lambda_{I_{\,i}}, \lambda_q, \lambda_s, \lambda_c$ (isospin, light, strange, and charm {\it quark fugacity factors}), and $\gamma_q,\gamma_s,\gamma_c$ (light, strange, and charm {\it quark phase space occupancies}) \cite{Torrieri:2004zz}. In this work we are not interested in the very small isospin effects and we neglect the contributions from the charmed hadrons. In this case we may write
\begin{eqnarray}
\Upsilon_i &=& \left(
\lambda_q \gamma_q\right)^{N^i_q}
\left(\lambda_s \gamma_s\right)^{N^i_s}
\left(\lambda_{\bar q} \gamma_{\bar q}\right)^{N^i_{\bar q}}
\left(\lambda_{\bar s} \gamma_{\bar s}\right)^{N^i_{\bar s}},
\label{upsilons}
\end{eqnarray}
where $\lambda_{q}=\lambda^{-1}_{\bar q}, \lambda_{s}=\lambda^{-1}_{\bar s}$, $\gamma_{q}=\gamma_{\bar q}$, and $\gamma_{s}=\gamma_{\bar s}$. In Eq.~(\ref{upsilons}), $N^i_q$ and $N^i_s$ are the numbers of light $(u,d)$ and strange $(s)$ quarks in the $i$th hadron, while $N^i_{\bar q}$ and $N^i_{\bar s}$  are the numbers of the antiquarks in the same hadron.

Equation (\ref{upsilons}) can be rewritten in terms of chemical
potentials using the Gell-Mann--Nishijima formulas. This leads to
the two relations: $\lambda_q = \exp(\mu_B/3T)$ and
\mbox{$\lambda_s = \exp((-3 \mu_S + \mu_B)/3T)$}, which finally
gives
\begin{eqnarray}
\Upsilon_i=
 \gamma_q^{N^i_q+N^i_{\bar q}} \gamma_s^{N^i_s+N^i_{\bar s}}  \exp \left( \frac{ \mu_B B_i  + \mu_S S_i}{T}\right).
\label{upsiNeq}
\end{eqnarray}
The equilibrium model may be treated as the special case of the
non-equilibrium model with $\gamma_{q}=\gamma_{s}=1$, see Refs.~\cite{Braun-Munzinger:1994xr,Braun-Munzinger:1995bp,Braun-Munzinger:1999qy,Braun-Munzinger:2001ip,Stachel:2013zma},
or with $\gamma_{q}=1$ and $\gamma_{s}<1$
\cite{Becattini:1997uf,Becattini:2000jw,Becattini:2003wp,Becattini:2012xb}.
In Eq.~(\ref{upsiNeq}) $B_i$ and $S_i$ are the baryon number
and strangeness of the $i$th particle, and $\mu$'s are the
corresponding chemical potentials.

In NEQ SHM we use the values from \cite{Petran:2013lja}. Since the
chemical potentials $\mu_B$ and $\mu_S$ found in
\cite{Petran:2013lja} are very small, we set them equal to zero,
which gives $\lambda_q=\lambda_s=1$. The other
parameters  depend on the centrality of the collision $c$. In
this work, we present our results for the two centrality classes:
$c$=0\%--5\% and $c$=30\%--40\%. The results for other centralities
are similar and will be published in a separate article.

For the most central collisions, $c$=0\%--5\%, we use
\begin{eqnarray}
T = 138.0\, \hbox{MeV},  \quad
\gamma_q = 1.63, \quad \gamma_s = 2.05, \quad
\label{NEQpar05}
\end{eqnarray}
and for semi-peripheral collisions, $c$=30\%--40\%, we take
\begin{eqnarray}
T = 139.85\, \hbox{MeV}, \quad
\gamma_q = 1.62, \quad \gamma_s = 2.0. \quad
\label{NEQpar3040}
\end{eqnarray}
The large value of $\gamma_q$ and low temperature mean that we deal with a supercooled QGP at the LHC, that is very close to the pion Bose-Einstein
condensation (BEC). The non-equilibrium chemical potential of pions, $\mu_{\pi}=2T\ln\gamma_q\simeq 134.9~$MeV, is very close to the mass of the
$\pi^0$ meson, $m_{\pi^0}\simeq 134.98$. Therefore, one should observe an enhancement in the production of pions at low values of $p_T$. In the EQ SHM version we use
\begin{eqnarray}
T = 165.6\, \hbox{MeV}. 
\label{EQpar}
\end{eqnarray}
The equilibrium value of the freeze-out temperature was used recently in \cite{Rybczynski:2012ed}. It is also very close to the value used in \cite{Kalweit:2012zz} to interpret the ALICE data.

Having fixed the values of the thermodynamic parameters, we have to define the remaining two geometric parameters. The Cracow model is boost-invariant and cylindrically symmetric \cite{Broniowski:2001we}.
The freeze-out hypersurface $\Sigma$ is defined by the equations
\begin{eqnarray}
t^2 = \tau_f^2 + x^2 + y^2 + z^2, \quad
x^2 + y^2 \leq r_{\rm max}^2,
\label{krk}
\end{eqnarray}
and the flow at freeze-out has the Hubble form, \mbox{$u^\mu = x^\mu/\tau_f$}.

\begin{figure}[t!]
\begin{center}
\includegraphics[angle=0,width=0.45\textwidth]{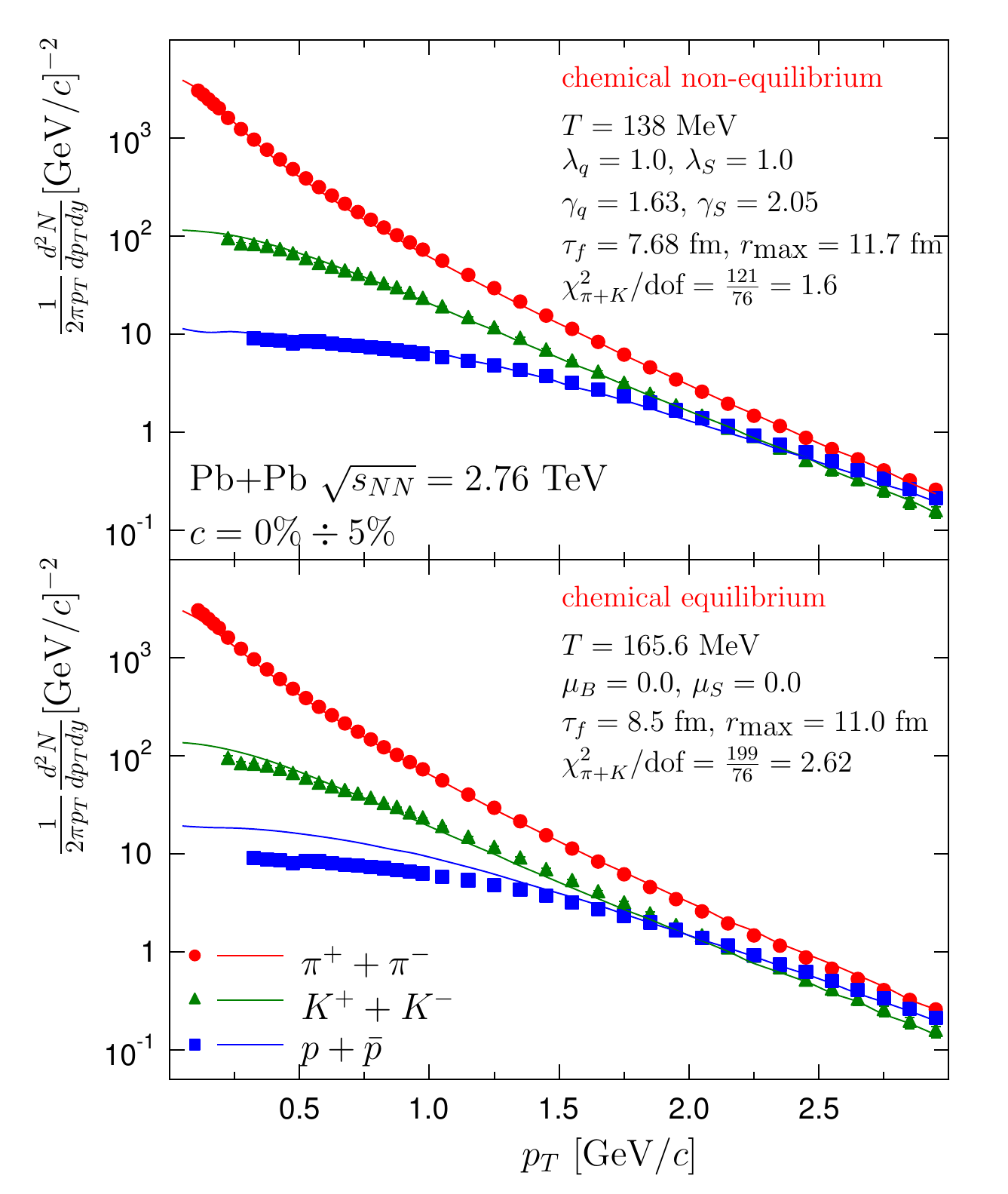}
\end{center}
 \vspace{-0.3cm}
 \caption{(Color online) Transverse-momentum spectra of pions (red circles), kaons (green triangles) and protons (blue squares)
 measured in Pb+Pb collisions at $\sqrt{s_{\rm NN}}$ = 2.76 TeV in the centrality class $c$=0\%--5\% \cite{Abelev:2012wca,Abelev:2013vea} compared to
 the model predictions (red, green, blue solid lines for pions, kaons, and protons, respectively). Upper panel shows the NEQ SHM
 predictions, while the lower panel shows the EQ SHM results.}
 \label{fig:fig1}
\end{figure}

\begin{figure}[t!]
\begin{center}
\includegraphics[angle=0,width=0.45\textwidth]{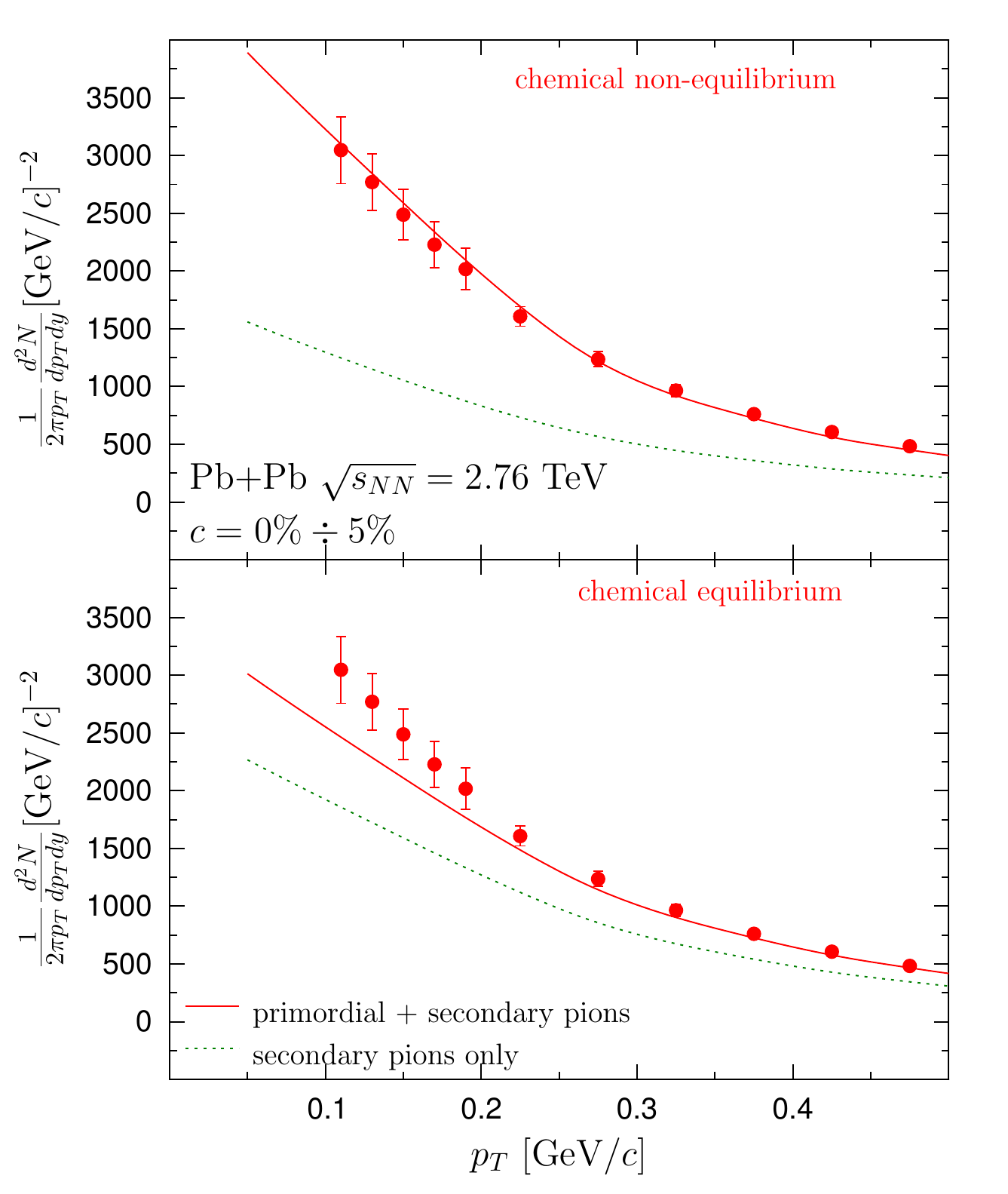}
\end{center}
 \vspace{-0.3cm}
 \caption{(Color online) Experimental and model transverse-momentum spectra of pions in the low-$p_T$ region.
 The theoretical predictions show secondary pions (dashed lines) and primary+secondary pions (solid lines). }
 \label{fig:fig2}
\end{figure}

We fix the geometric parameters by fitting the pion and kaon spectra. In NEQ SHM this gives:
\begin{eqnarray}
\tau_f = 7.68\,\hbox{fm}, \quad r_{\rm max} = 11.7\,\hbox{fm}, \quad \hbox{for $c$=0\%--5\%}, \nonumber \\
\tau_f = 4.83\,\hbox{fm}, \quad r_{\rm max} = 7.5\,\hbox{fm}, \quad \hbox{for $c$=30\%--40\%},
\label{krkI}
\end{eqnarray}
while in EQ SHM we find
\begin{eqnarray}
\tau_f = 8.5\,\hbox{fm}, \quad r_{\rm max} = 11.0 \,\hbox{fm}, \quad \hbox{for $c$=0\%--5\%}, \nonumber \\
\tau_f = 5.5\,\hbox{fm}, \quad r_{\rm max} = 7.0\,\hbox{fm}, \quad \hbox{for $c$=30\%-40\%}.
\label{krkII}
\end{eqnarray}

\begin{figure}[t]
\begin{center}
\includegraphics[angle=0,width=0.45\textwidth]{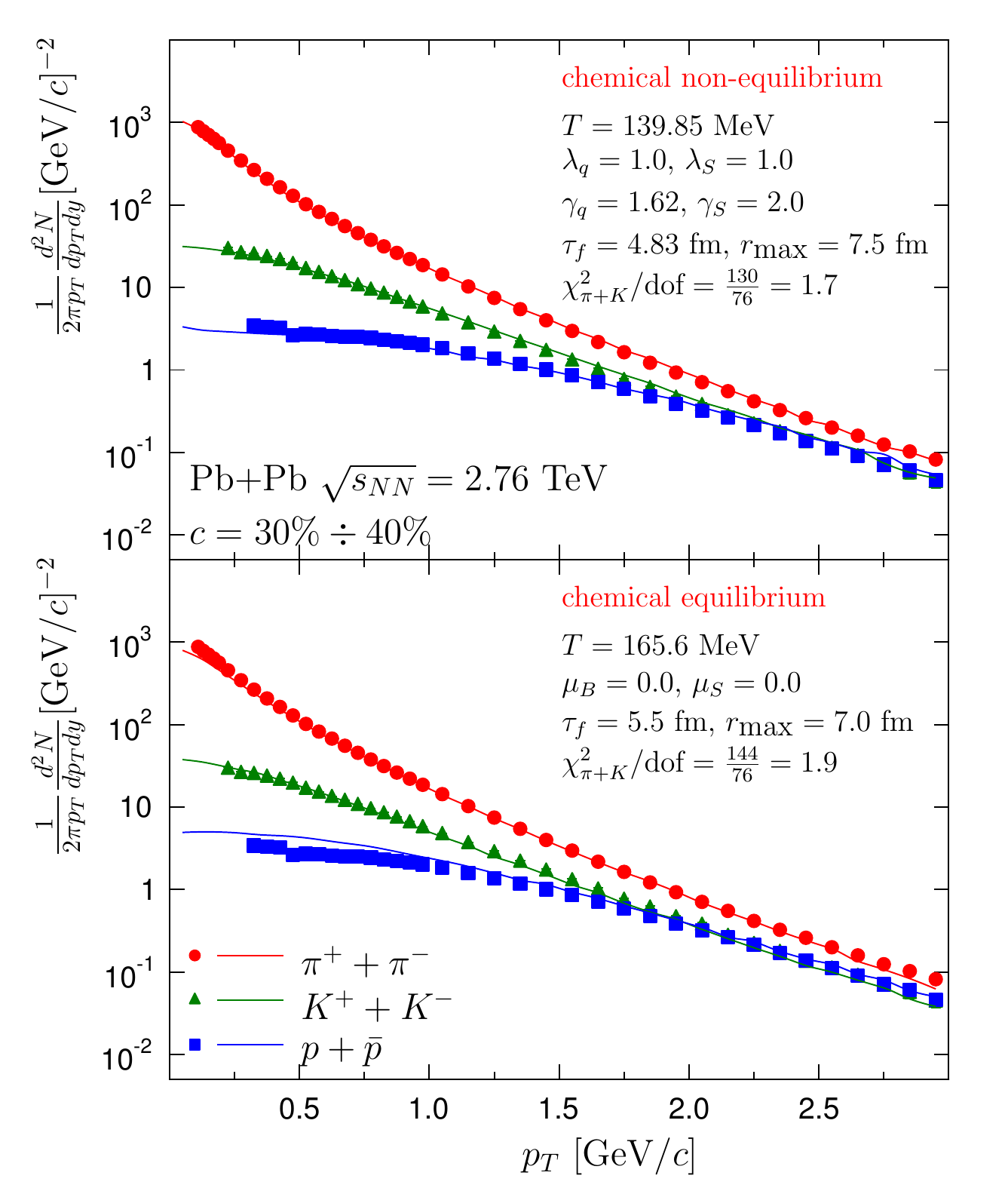}
\end{center}
 \vspace{-0.3cm}
 \caption{(Color online) The same as Fig.~1 but for the centrality class $c$=30\%--40\%.}
\label{fig:fig3}
\end{figure}

\begin{figure}[t]
\begin{center}
\includegraphics[angle=0,width=0.45\textwidth]{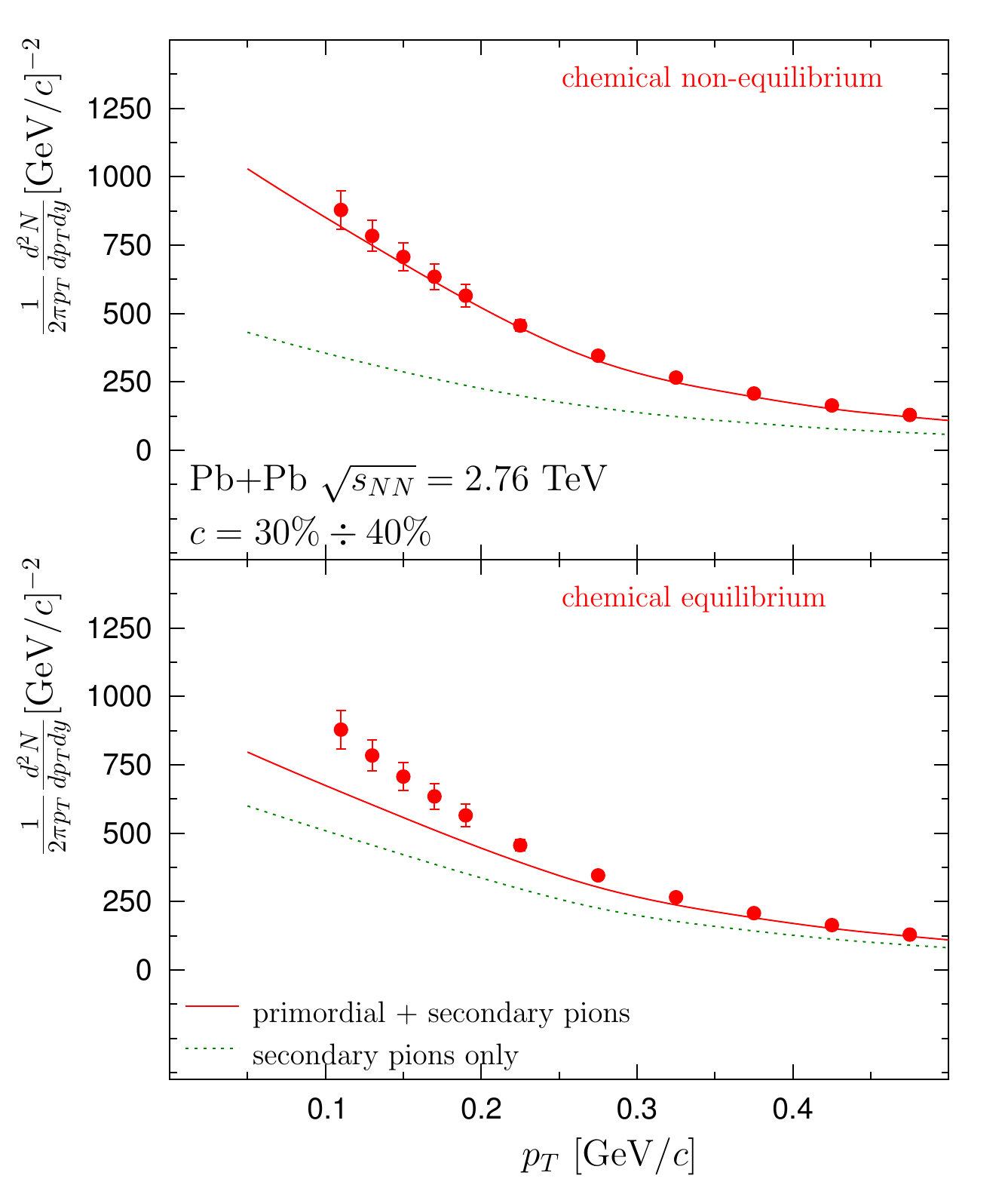}
\end{center}
 \vspace{-0.3cm}
 \caption{(Color online) The same as Fig.~2 but for $c$=30\%--40\%.}
 \label{fig:fig4}
\end{figure}

In Fig.~1 we show our results for the transverse-momentum spectra
of pions ($\pi^+$ and $\pi^-$), kaons ($K^+$ and $K^-$), and
protons ($p$ and ${\bar p}$) in the range up to 3~GeV/c for the
centrality class $c$=0\%--5\%. The model calculations are compared
with the data taken from \cite{Abelev:2012wca,Abelev:2013vea}. The
upper part describes the comparison of the experimental results
with the predictions of the chemical non-equilibrium model, and
the lower part describes the comparison with the chemical
equilibrium model.

The model calculations have been corrected for the weak decays.
The fit is based on the $\chi^2$ method applied to pions and kaons
only, with the errors taken from
\cite{Abelev:2012wca,Abelev:2013vea}. The values of $\chi^2$
divided by the number of degrees of freedom are shown in the
plots. They indicate the good quality of the fits (with advantage
of NEQ SHM).

Since protons are not included in the fitting procedure, their
spectra may be treated as predictions of the model. From the upper
part of Fig.~1 we conclude that the chemical non-equilibrium model
describes well the proton spectrum, in addition to the proton
yield described earlier in \cite{Petran:2013lja}. The lower part
of Fig.~1 displays a problem with the correct description of the
proton yield within EQ SHM.

In Fig.~2 we show in greater detail the low-$p_T$ spectra ($p_T <$
500 MeV/c) of pions for the centrality class \mbox{$c$=0\%--5\%}.
The upper (lower) part gives the comparison with NEQ SHM (EQ SHM)
model. The dashed lines describe the spectra of  secondary pions produced by strong resonance decays, whereas the solid lines describe the full spectra. The difference between the solid and
dashed lines describes the contribution from the directly produced pions.

One can observe a remarkable agreement between the model predictions and the data in the case of chemical non-equilibrium. The very good description of the data may be attributed to the use of the $\gamma_q$ quark occupancy factors, which are significantly larger than unity. These factors reduce denominators of the pion distribution functions and, consequently, lead to a very much peaked behavior of the pion primordial distributions at very small
values of $p_T$. By comparison of the upper and lower parts, we
can observe that such a behavior is absent in the equilibrium
model.

In Fig.~3 we show the pion, kaon, and proton spectra for the
centrality class $c$=30\%--40\%. The presentation of the results is
analogous to that used in Fig.~1. Similarly, Fig.~4 shows the
low-$p_T$ region of the pion spectrum for $c$=30\%--40\%. The
qualitative and quantitative features of the model calculations
are the same as those found in our study of the most central
collisions. We have also studied other centrality classes and
found the same type of behavior. The results obtained for the
complete set of centralities will be published in a forthcoming
publication.

In conclusions, we emphasize that the chemical non-equilibrium version of the thermal model combined with the single-freeze-out scenario explains very well the transverse-momentum spectra of pions, kaons, and protons. This approach eliminates the proton anomaly and simultaneously explains the low-transverse-momentum enhancement of pions.

Correct description of the low-transverse-momentum enhancement of pions within NEQ SHM suggests that it may be interpreted as a signature of the onset of pion condensation in ultra-relativistic
heavy-ion collisions at the LHC energies. This phenomenon may be connected with a recent experimental finding of a coherent
contribution to the pion production at the LHC
\cite{Abelev:2013pqa}. In the past, there were many suggestions to
look for pion condensation, see, for example,
\cite{Begun:2006gj,Begun:2008hq} and references therein. Very recently, the possibility of the gluon condensation has been suggested in \cite{Blaizot:2011xf,Blaizot:2013lga} in the context of thermalization of the quark-gluon plasma. This might have also a connection with the present work. 

Definitely, the low-$p_T$ enhancement of pions at the LHC deserves further theoretical and experimental studies. In particular, our results suggest that it would be very interesting to measure the pion spectrum at smaller values of $p_T$ than those available at the moment.

We thank Christina Markert and Jinfeng Liao for discussions. V.B. and W.F. were supported in part by the Polish National Science Center grant with decision No. DEC-2012/06/A/ST2/00390.


\begin{thebibliography}{20}

\bibitem{Koch:1985hk}
P.~Koch, J.~Rafelski, South Afr. J. Phys. {\bf 9}, 8 (1986).

\bibitem{Cleymans:1992zc}
J.~Cleymans, H.~Satz, Z. Phys. C {\bf 57}, 135 (1993).

\bibitem{Cleymans:1998fq}
J.~Cleymans, K.~Redlich, Phys. Rev. Lett. {\bf 81}, 5284 (1998).

\bibitem{Gazdzicki:1998vd}
M.~Gazdzicki, M.~I. Gorenstein, Acta Phys. Polon. B {\bf 30}, 2705 (1999).

\bibitem{Braun-Munzinger:1994xr}
P.~Braun-Munzinger, J.~Stachel, J.~P. Wessels, N.~Xu, Phys. Lett. B {\bf 344}, 43 (1995).

\bibitem{Cleymans:1996cd}
J.~Cleymans, D.~Elliott, H.~Satz, R.~L. Thews, Z. Phys. C {\bf 74}, 319 (1997).

\bibitem{Becattini:2000jw}
F.~Becattini, J.~Cleymans, A.~Keranen, E.~Suhonen, K.~Redlich, Phys. Rev. C {\bf 64}, 024901 (2001).

\bibitem{Sollfrank:1993wn}
J.~Sollfrank, M.~Gazdzicki, U.~W. Heinz, J.~Rafelski, Z. Phys. C {\bf 61}, 659
  (1994).

\bibitem{Schnedermann:1993ws}
E.~Schnedermann, J.~Sollfrank, U.~W. Heinz, Phys. Rev. C {\bf 48}, 2462 (1993).

\bibitem{Braun-Munzinger:1995bp}
P.~Braun-Munzinger, J.~Stachel, J.~P. Wessels, N.~Xu, Phys. Lett. B {\bf 365}, 1 (1996).

\bibitem{Becattini:1997uf}
F.~Becattini, J. Phys. G {\bf 23}, 1933 (1997).

\bibitem{Yen:1998pa}
G.~D. Yen, M.~I. Gorenstein, Phys. Rev. C {\bf 59}, 2788 (1999).

\bibitem{Braun-Munzinger:1999qy}
P.~Braun-Munzinger, I.~Heppe, J.~Stachel, Phys. Lett. B {\bf 465}, 15 (1999).

\bibitem{Becattini:2003wp}
F.~Becattini, M.~Gazdzicki, A.~Keranen, J.~Manninen, R.~Stock, Phys. Rev. C {\bf
  69}, 024905 (2004).

\bibitem{Braun-Munzinger:2001ip}
P.~Braun-Munzinger, D.~Magestro, K.~Redlich, J.~Stachel, Phys. Lett. B {\bf 518}, 41 (2001).

\bibitem{Florkowski:2001fp}
W.~Florkowski, W.~Broniowski, M.~Michalec, Acta Phys. Polon. B {\bf 33}, 761 (2002).

\bibitem{Broniowski:2001we}
W.~Broniowski, W.~Florkowski, Phys. Rev. Lett. {\bf 87}, 272302 (2001).

\bibitem{Broniowski:2001uk}
W.~Broniowski, W.~Florkowski, Phys. Rev. C {\bf 65}, 064905 (2002).

\bibitem{Retiere:2003kf}
F.~Retiere, M.~A. Lisa, Phys. Rev. C {\bf 70}, 044907 (2004).


\bibitem{Abelev:2012wca}
  B.~Abelev {\it et al.}  [ALICE Collaboration],
  Phys.\ Rev.\ Lett.\  {\bf 109}, 252301 (2012).

\bibitem{Abelev:2013vea}
  B.~Abelev {\it et al.}  [ALICE Collaboration],
    Phys.\ Rev.\ C {\bf 88}, 044910 (2013).

 \bibitem{Kalweit:2012zz}
  A.~Kalweit [ALICE Collaboration],
   Acta Phys.\ Polon.\ Supp.\  {\bf 5}, 225 (2012).

\bibitem{Stachel:2013zma}
  J.~Stachel, A.~Andronic, P.~Braun-Munzinger and K.~Redlich,  arXiv:1311.4662 [nucl-th].

\bibitem{Becattini:2012xb}
  F.~Becattini, M.~Bleicher, T.~Kollegger, T.~Schuster, J.~Steinheimer and R.~Stock,
    Phys.\ Rev.\ Lett.\  {\bf 111}, 082302 (2013).

\bibitem{Petran:2013qla}   M.~Petran and J.~Rafelski, Phys.\ Rev.\ C {\bf 88}, 021901 (2013).

\bibitem{Petran:2013lja}
  M.~Petran, J.~Letessier, V.~Petracek and J.~Rafelski,    Phys.\ Rev.\ C {\bf 88}, 034907 (2013).

\bibitem{Song:2013qma}
  H.~Song, S.~Bass and U.~W.~Heinz,
  arXiv:1311.0157 [nucl-th].
  
\bibitem{Pang:2013pma} 
  L.~Pang, Q.~Wang and X.~-N.~Wang,
  arXiv:1309.6735 [nucl-th].

\bibitem{Prorok:2006ve}
D.~Prorok, Phys.\ Rev.\ C {\bf 75}, 014903 (2007).

\bibitem{Rafelski:2011ek}
  J.~Rafelski,    Acta Phys.\ Polon.\ B {\bf 43}, 829 (2012).
  
\bibitem{Floris} Michele Floris (ALICE Collaboration) review talk presented at the Quark Matter 2014 conference, Darmstadt, Germany.

\bibitem{Kisiel:2005hn}
  A.~Kisiel, T.~Taluc, W.~Broniowski and W.~Florkowski,  Comput.\ Phys.\ Commun.\  {\bf 174}, 669 (2006).

  \bibitem{Chojnacki:2011hb}
  M.~Chojnacki, A.~Kisiel, W.~Florkowski and W.~Broniowski,  Comput.\ Phys.\ Commun.\  {\bf 183}, 746 (2012).

 \bibitem{Torrieri:2004zz}
  G.~Torrieri, S.~Steinke, W.~Broniowski, W.~Florkowski, J.~Letessier and J.~Rafelski,
    Comput.\ Phys.\ Commun.\  {\bf 167}, 229 (2005).

   \bibitem{Rybczynski:2012ed}
  M.~Rybczynski, W.~Florkowski and W.~Broniowski,
    Phys.\ Rev.\ C {\bf 85}, 054907 (2012).

   \bibitem{Abelev:2013pqa}
  B.~B.~Abelev {\it et al.}  [ALICE Collaboration], arXiv:1310.7808 [nucl-ex].


    \bibitem{Begun:2006gj}
  V.~V.~Begun and M.~I.~Gorenstein,
    Phys.\ Lett.\ B {\bf 653}, 190 (2007).

    \bibitem{Begun:2008hq}
  V.~V.~Begun and M.~I.~Gorenstein,
    Phys.\ Rev.\ C {\bf 77}, 064903 (2008).

\bibitem{Blaizot:2011xf}
  J.~-P.~Blaizot, F.~Gelis, J.~-F.~Liao, L.~McLerran and R.~Venugopalan,
   Nucl.\ Phys.\ A {\bf 873}, 68 (2012).

   \bibitem{Blaizot:2013lga}
   J.~-P.~Blaizot, J.~Liao and L.~McLerran,
  Nucl.\ Phys.\ A {\bf 920}, 58 (2013).

\end{thebibliography}
\end{document}